\documentclass[twocolumn,final,aps,prb,showpacs,superscriptaddress,amsmath,amssymb,amsfonts,floatfix]{revtex4-2}
\usepackage{epsfig}
\usepackage[dvipsnames,usenames]{color}
\usepackage{xcolor}
\usepackage{tcolorbox}
\usepackage{amsmath}
\usepackage{comment}
\usepackage{array}
\usepackage{multirow}
\usepackage{tabularx}
\usepackage{float}
\usepackage[colorlinks=true,bookmarks=true, pdfborder={0 0 0},linkcolor=blue,urlcolor=blue,     breaklinks=true,bookmarksnumbered=true]{hyperref}
\usepackage[normalem]{ulem}
\usepackage{subfigure}
\usepackage{orcidlink}
\emergencystretch=\maxdimen
\begin{document}

\title{Enhanced $s^\pm$-Wave Superconductivity in Electron-Doped La$_3$Ni$_2$O$_7$}

\author{Xun Liu\,\orcidlink{0009-0000-4906-3078}}
\affiliation{Institute for Quantum Science, School of Physical Science and Technology, \\ Soochow University, Suzhou 215006, China}

\author{Chao Deng\,\orcidlink{0009-0003-8433-2909}}
\affiliation{School of Physics, Northwest University, Xi’an 710127, China}

\author{Wenfeng Wu\,\orcidlink{0000-0002-6575-5813}}
\affiliation{Institute of Solid State Physics, TU Wien, 1040 Vienna, Austria}

\author{Liang Si\,\orcidlink{0000-0003-4709-6882}}
\email{siliang@nwu.edu.cn}
\affiliation{School of Physics, Northwest University, Xi’an 710127, China}
\affiliation{Shaanxi Key Laboratory for Theoretical Physics Frontiers, Xi'an 710127, China}
\affiliation{Fundamental Discipline Research Center for  Quantum Science and technology of Shaanxi Province, Xi'an 710127, China}

\author{Mi Jiang\,\orcidlink{0000-0002-9500-202X}}
\email{jiangmi@suda.edu.cn}
\affiliation{Institute for Quantum Science, School of Physical Science and Technology, \\ Soochow University, Suzhou 215006, China}
\affiliation{State Key Laboratory of Surface Physics and Department of Physics, Fudan University, Shanghai 200433, P. R. China}

\begin{abstract}

In cuprates, electron doping yields a much lower superconducting $T_c$ than hole doping. For recently discovered nickelate superconductors, the analogous doping strategies become more challenging. Consequently, while hole-doped Ruddlesden-Popper (RP) nickelates have been extensively studied, electron-doped RP nickelates remain rarely explored both experimentally and theoretically. Here we fill this gap by systematically investigating the two-orbital bilayer model for three representative systems: bulk La$_3$Ni$_2$O$_7$ at ambient pressure and 15\,GPa, and a heterostructure La$_3$Ni$_2$O$_7$:La$_3$Al$_2$O$_7$ that provides a feasible experimental route to electron doping. Using first-principle calculations and large-scale dynamical cluster quantum Monte Carlo simulations, we find that electron doping generically enhances $s^\pm$-wave pairing superconductivity (SC) in all three cases, with the heterostructure showing the highest $T_c$ in the underdoped regime. Furthermore, our results suggest an inter-orbital cooperative mechanism that the pairing on the $d_{x^2-y^2}$ orbital, induced by that on the $d_{z^2}$ orbital, plays a vital role in the SC. This work provides the theoretical prediction of enhanced SC in electron-doped RP nickelates and calls for future experimental verification.

\end{abstract}

\maketitle

%%%%%%%%%%%%%%%%%
%\section{Introduction}

\textbf{\textit{Introduction.}}
The discovery of high-temperature superconductivity (SC) in cuprates~\cite{cuprate1,cuprate2,cuprate3,cuprate4} has long been a central theme in condensed matter physics. A hallmark of these layered oxides is the strong electron–hole asymmetry in their phase diagrams: under hole doping, a broad superconducting dome emerges with $T_c$ reaching values above 100\,K \cite{cuprate5,cuprate6,cuprate7}, whereas electron doping yields a much narrower superconducting region and a significantly suppressed $T_c$~\cite{Nd3+}. This asymmetry mostly originates from the different electronic structures of hole- and electron-doped Fermi surfaces. 
Technically, in hole-doped cuprates, substitution of trivalent $A$-site cations (La$^{3+}$, Nd$^{3+}$ or Pr$^{3+}$) by divalent cations (e.g., Ca$^{2+}$, Sr$^{2+}$ or Ba$^{2+}$) introduces holes into the CuO$_2$ planes in cuprates such as La$_2$CuO$_4$ \cite{anderson1987resonating,attfield1998cation}.
Electron doping, on the other hand, is typically realized by replacing trivalent rare-earth ions (La$^{3+}$, Nd$^{3+}$ or Pr$^{2+}$) with tetravalent Ce$^{4+}$ (e.g., in Nd$_{2-x}$Ce$_x$CuO$_4$) \cite{Ce4+,Nd3+} or by substituting divalent $A$-site cations with trivalent ions (e.g., Nd$^{3+}$ or La$^{3+}$ in place of Sr$^{2+}$ or Ca$^{2+}$) \cite{smith1991electron,PhysRevMaterials.3.064803}.

\begin{figure}
\psfig{figure=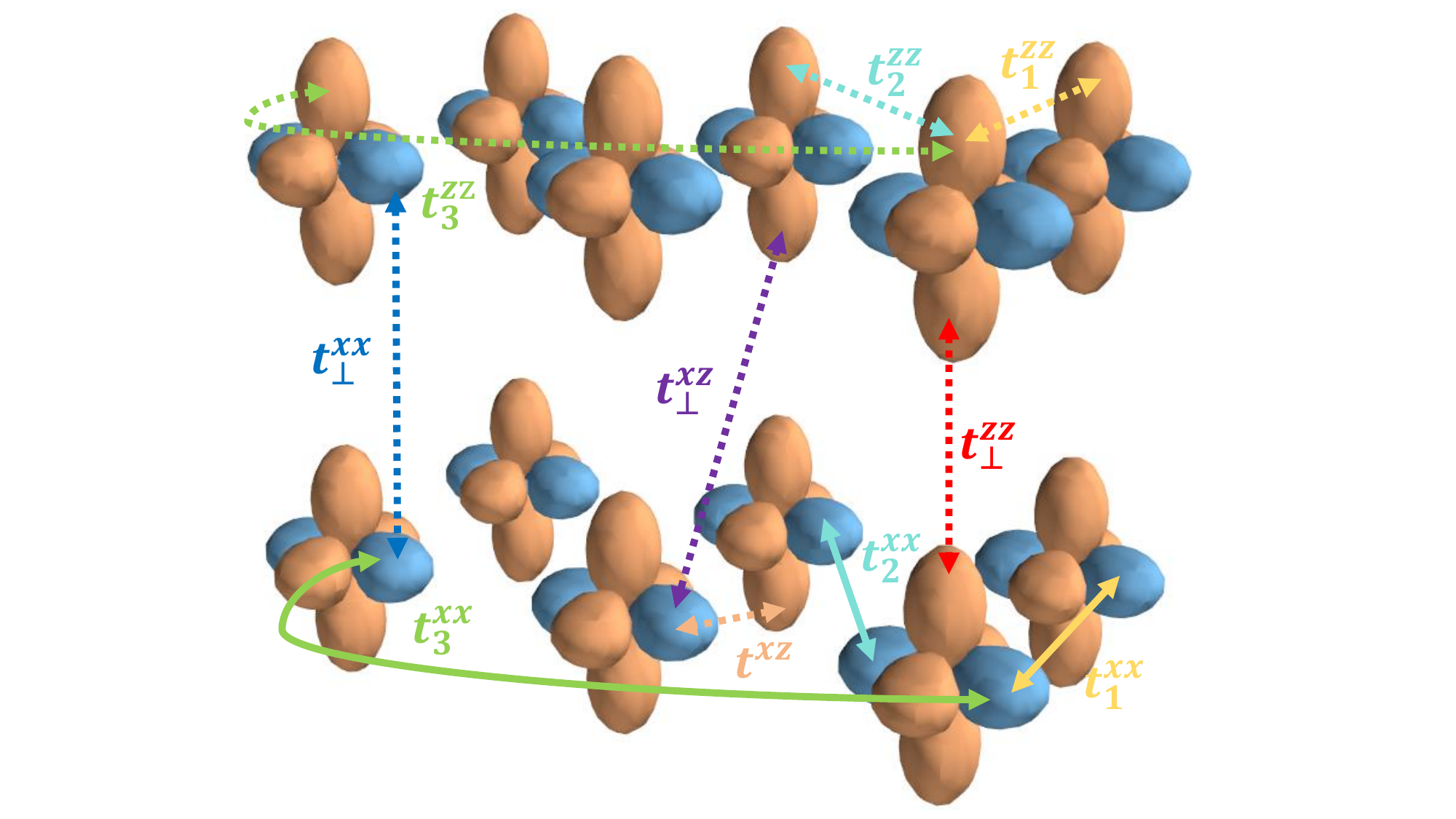,width=0.46\textwidth,clip=true,viewport=140 0 820 550}
\caption{Schematic illustration of the bilayer two-orbital model on a square lattice. The model incorporates three types of hoppings: (1) intralayer intra-orbital hoppings: nearest-neighbor ($t_1^{xx/zz}$), next-nearest-neighbor ($t_2^{xx/zz}$), and next-next-nearest-neighbor ($t_3^{xx/zz}$); (2) intralayer inter-orbital hopping ($t^{xz}$); and (3) interlayer hoppings: intra-orbital ($t_\perp^{xx/zz}$) and inter-orbital ($t_\perp^{xz}$). The values of these parameters for 0\,GPa, 15\,GPa and LaNi(Al)O are listed in Table. S1 in Sec.~I of Supplementary Materials (SM) \cite{SM}.}
\label{lattice}
\end{figure}

In sharp contrast to cuprates, the recently discovered nickelate superconductors, including the infinite-layer family $R$NiO$_2$ ($R$=La, Nd or Pr)~\cite{Ni1121,Ni1122,Ni1123} and the Ruddlesden-Popper (RP) phases La$_3$Ni$_2$O$_7$ \cite{WM2023,YHQ2024,PRX2024,JMST2024,CJG2024} and La$_4$Ni$_3$O$_{10}$ \cite{43101,43102,43103}, present a very different doping landscape. The nickel's trivalent (Ni$^{3+}$) or divalent (Ni$^{2+}$) oxidation states make it difficult to directly transplant the electron-doping schemes developed for cuprates. Few elements (such as Ce) can reliably adopt a stable $+4$ oxidation state under the relevant synthesis conditions, however, the electrons from Ce-$4f$ orbital lie energetically much deeper (by several eV) than the Ni-3$d$ electrons, hence Ce substitution does not effectively transfer/dope electrons into the $e_g$ orbital in Ni$^2+$ or Ni$^+$. Consequently, the recently reported experimental studies on RP phase nickelates have so far been focused on samples with hole doping, typically by partially substituting La$^{3+}$ with Sr$^{3+}$ to simultaneously stabilize the lattice structure and introduce holes into the NiO$_2$ planes  \cite{Sr327,NSR,halfdome}. Electron-doped RP nickelates remain almost completely unexplored experimentally and rarely discussed theoretically.

\begin{figure}
\psfig{figure=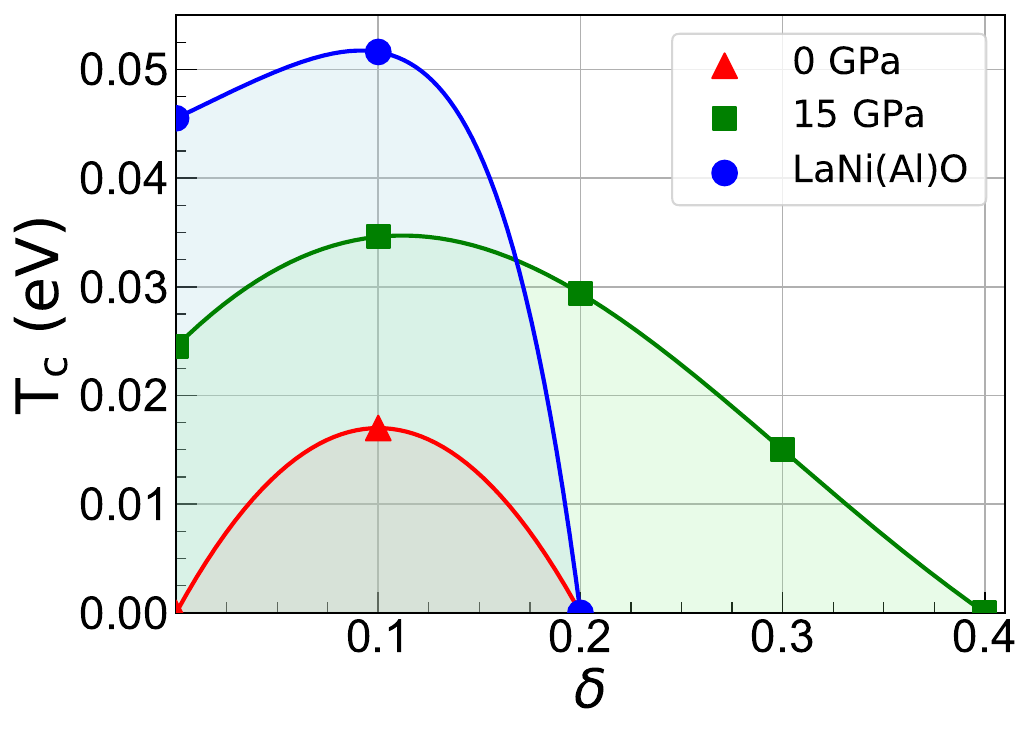,width=0.48\textwidth,clip}
\caption{
Extrapolated $T_c$ as a function of electron doping level $\delta$ for three different cases with the $N_c$=4$\times$2 cluster.}
\label{phase}
\end{figure}

Given that electron doping in cuprates already yields unconventional SC (albeit with lower $T_c$) and that the nickelate's electronic structure shares important similarities (and also differences such as the interstitial-$s$ orbital \cite{PhysRevB.100.205138,gu2020substantial,xqm6-wr7n} and pocket bands) with that of cuprates (e.g., quasi-two-dimensional $d_{x^2-y^2}$-derived Fermi surfaces), it becomes imperative and essential to raise the question: can electron doping also induce or even enhance SC in RP nickelates ? To answer this question, we need not only a viable experimental route to electron doping, but also a reliable (many-body level) theoretical framework to compute the superconducting properties beyond conventional mean-field approximations such as density-functional theory (DFT).

In this work, we perform large-scale many-body calculation toward filling this gap. We construct a (minimal) two-orbital bilayer tight-binding model that captures the low-energy physics of three distinct systems: (i) bulk La$_3$Ni$_2$O$_7$ at ambient pressure (0\,GPa), (ii) the same compound under 15\,GPa where SC has been observed experimentally \cite{WM2023}, and (iii) a heterostructure La$_3$Ni$_2$O$_7$:La$_3$Al$_2$O$_7$ (LaNi(Al)O)~\cite{SL}. The latter is designed to achieve electron doping via charge transfer from the insulating La$_3$Al$_2$O$_7$ spacer layers. Using first-principles density functional theory (DFT)~\cite{DFT1,DFT2,DFT3,DFT4,DFT5,DFT6,DFT7,DFT8} to obtain accurate hopping parameters and employing large‑scale dynamical cluster quantum Monte Carlo (QMC) simulations to treat strong correlations, we systematically study the pairing tendencies in these electron‑doped systems. We find that electron doping universally enhances $s^\pm$-wave pairing SC in all three cases, with the heterostructured LaNi(Al)O exhibiting the highest $T_c$ in the underdoped regime. Moreover, our results reveal an inter‑orbital cooperative mechanism: the $s^\pm$-wave pairing on the 3$d_{x^2-y^2}$  orbital appears to be induced by that on the Ni-3$d_{z^2}$ orbital. This work not only provides the theoretical prediction of enhanced SC in electron-doped RP nickelates but also proposes a concrete and experimentally feasible heterostructure design to realize it, stimulating future experimental efforts to explore the rich and largely uncharted electron-doped side of nickelate superconductors.

The electronic ground state of the parent compound La$_3$Ni$_2$O$_7$ remains under debate. While stoichiometric samples have been reported to exhibit a correlated metallic state \cite{kakoi2024multiband,PhysRevB.83.245128}, semiconducting transport has also been observed \cite{hou2023emergence}, and oxygen nonstoichiometry, such as oxygen vacancies, can drive the system toward an insulating state \cite{dong2024visualization,wang2025mottness}. In our calculations, we adopt the pristine, high-symmetry $I4/mmm$ phase  \cite{wang2024structure} to extract the model parameters, consistent with the pressurized superconducting phase of La$_3$Ni$_2$O$_7$.

%\section{Model}

\begin{figure*}
\psfig{figure=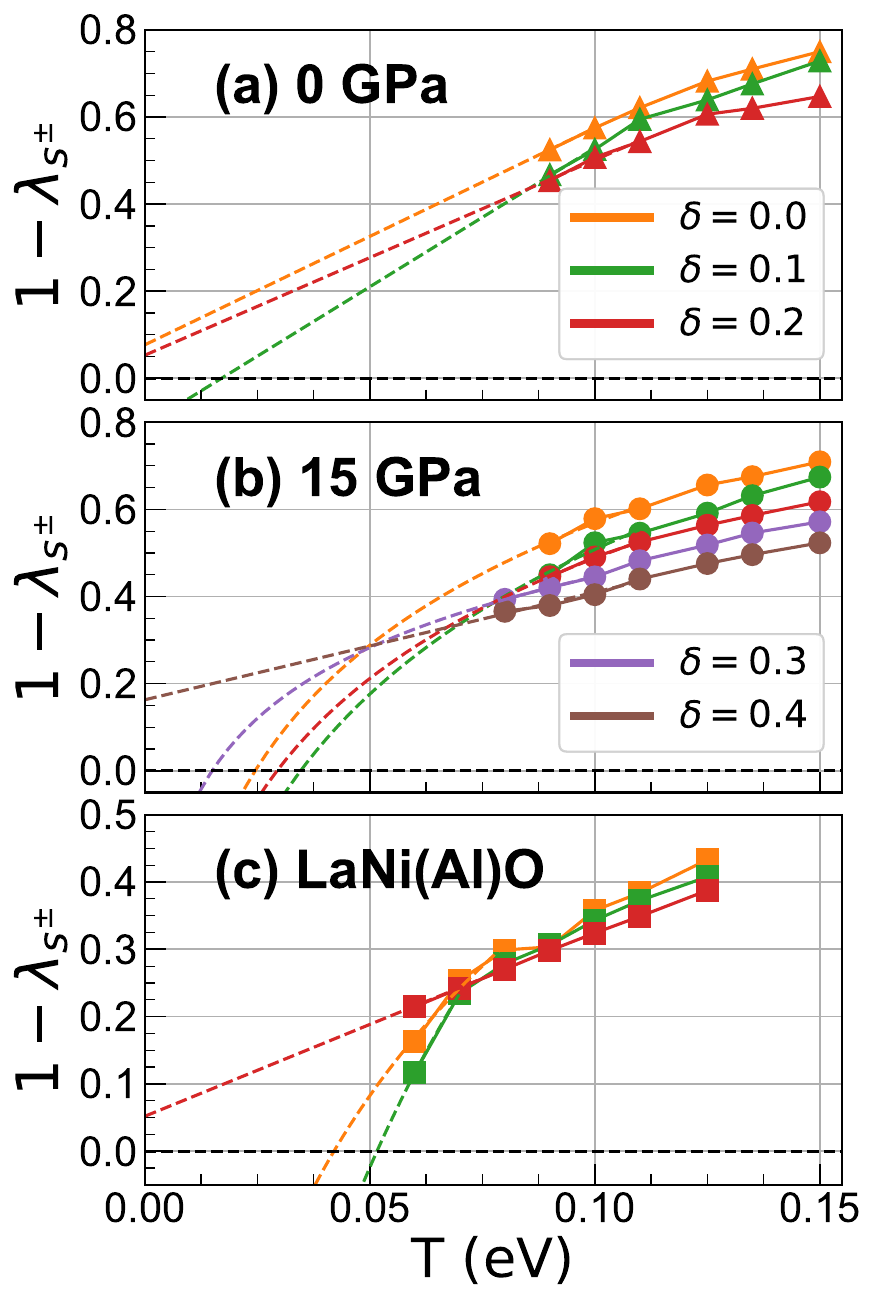,width=0.32\textwidth,clip}
\raisebox{0.5cm}{\psfig{figure=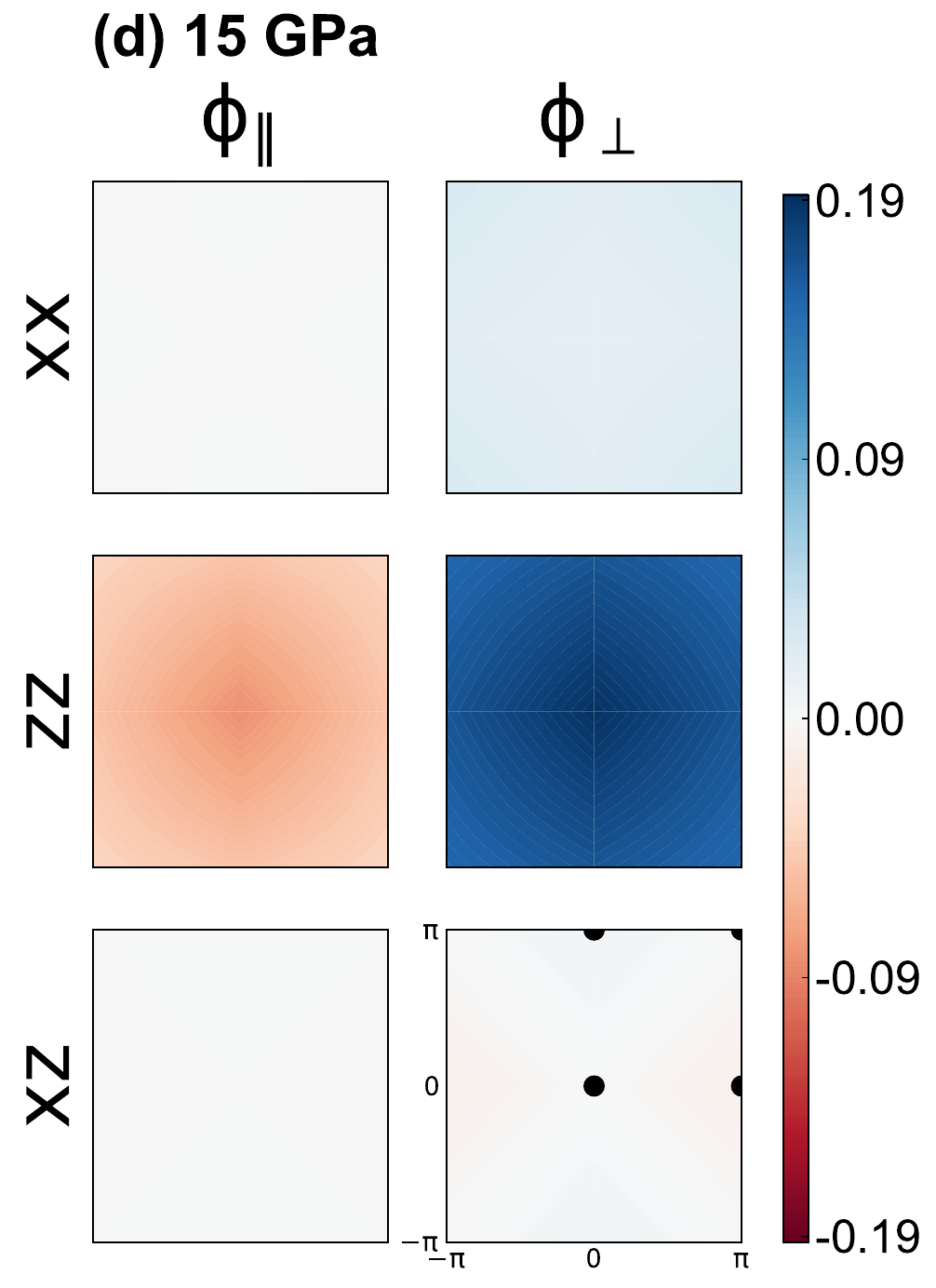,width=0.32\textwidth,clip}}
\raisebox{0.5cm}{\psfig{figure=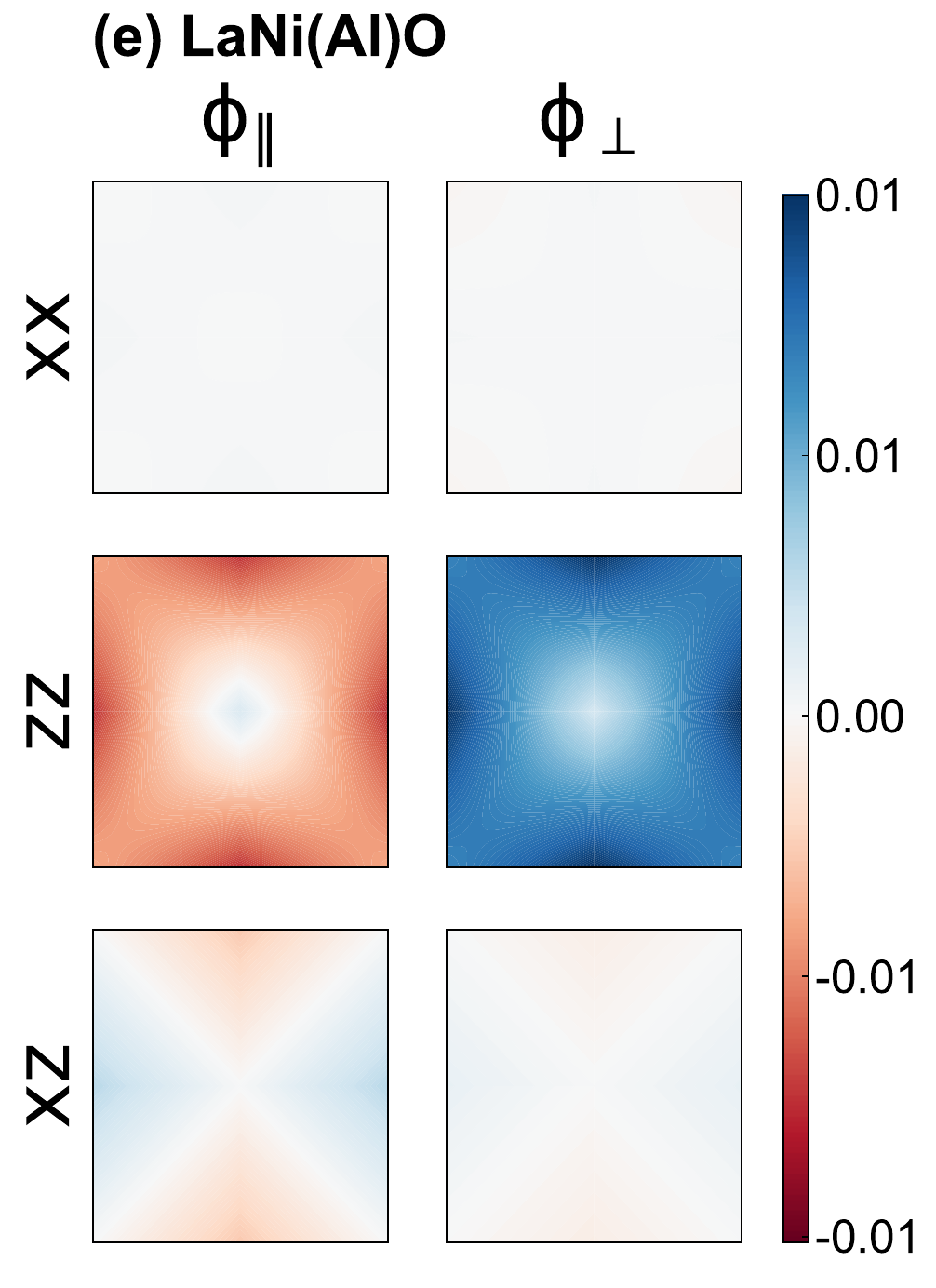,width=0.32\textwidth,clip}}
\caption{
Temperature dependence of the eigenvalue 1-$\lambda_{s^\pm}$ for extrapolating $T_c$ under three distinct conditions: (a) 0\,GPa (triangles, $\delta$= 0.0, 0.1, 0.2), (b) 15\,GPa (circles, $\delta$=0.0, 0.1, 0.2, 0.3, 0.4), and (c) LaNi(Al)O (squares, $\delta$=0.0, 0.1, 0.2). The corresponding BSE eigenvectors, highlighting stronger interlayer pairing, are shown for two representative cases: (d) 15\,GPa with $\delta$=0.3 at $T$=0.09\,eV, and (e) LaNi(Al)O with $\delta = 0.1$ at $T$=0.06\,eV. Here, $\phi_\|$ and $\phi_\perp$ denote the intralayer and interlayer components, respectively. The high-symmetry $\mathrm{K}$ points included in $N_c$=4$\times$2 cluster are marked by black circles in (d).}
\label{BSE}
\end{figure*}

\textbf{\textit{Model.}} We consider the bilayer two-orbital model ~\cite{YDX} for the bulk La$_3$Ni$_2$O$_7$ on a two-dimensional square lattice as shown in Fig.~\ref{lattice} with the Hamiltonian
\begin{align}
    H& =  H_0+H_U \notag \\
    H_0 &= \sum_{ i j \sigma m l\nu} t_\nu^{xx/zz} c_{i ml \sigma}^\dagger c_{j ml \sigma}+t^{xz}\sum_{\langle ij \rangle \sigma mll'}c_{i ml \sigma}^\dagger c_{j ml' \sigma} \notag \\
    &\quad+ t_\perp^{xz} \sum_{\langle ij \rangle ll'\sigma} c_{i 1l \sigma}^\dagger c_{j 2nl' \sigma}+t_\perp^{xx/zz} \sum_{i l\sigma} c_{i 1l \sigma}^\dagger c_{i 2l\sigma}\notag \\
    &\quad + (\epsilon^{x/z} - \mu)\sum_{iml\sigma} n_{i ml \sigma} + \mathrm{H}.c. \notag \\
    H_U &= U \sum_{i ml} n_{i ml \uparrow} n_{i ml \downarrow} + U' \sum_{imll'\sigma\sigma'} n_{i ml\sigma} n_{i ml' \sigma'} \notag \\
    &\quad+ (U'-J) \sum_{imll'\sigma} n_{i ml\sigma} n_{i ml' \sigma} \notag
\end{align}
where $c_{i\sigma}^{\dagger}$ ($c_{i\sigma}$) creates (annihilates) an electron at site $i$ in layer $m$=1, 2 with spin $\sigma$=$\uparrow$, $\downarrow$ and orbital $l$=$d_{x^2-y^2}$ ($x$) or $d_{z^2}$ ($z$). The hoppings $t_{\nu}^{xx/zz}$ include nearest-, next-nearest-, and next-next-nearest-neighbor terms $t_1^{xx/zz}$, $t_2^{xx/zz}$, $t_3^{xx/zz}$ for the two orbitals. $t_{\perp}^{xx/zz}$ are interlayer intra-orbital hoppings, and $t^{xz}$, $t_{\perp}^{xz}$ are intralayer and interlayer inter-orbital hoppings. 
The details on the tight-binding model used to describe superconductivity in La$_3$Ni$_3$O$_7$, the corresponding parameters (derived from DFT calculations and Wannier projections), and the (non-interacting) DFT band structures are shown in Table~S1 and Fig.~S1 in Sec.~I of SM \cite{SM}.
The on-site Coulomb interaction is $U$=4.0\,eV, inter-orbital $U'$=
$U$-2$J$=2.4\,eV with Hund's coupling $J$=$U$/5=0.8\,eV. The chemical potential $\mu$ tunes the average total density $n$ per bilayer $e_g$ orbitals, with electron-doping level $\delta$=$n$-3.0, and the density distribution is displayed in Fig.~S2 in Sec.~II of SM~\cite{SM}. The model is solved using dynamical cluster approximation (DCA) with CT-QMC cluster solver \cite{Hettler98,Maier05,code,GullCTAUX}. The superconducting properties are obtained by solving the Bethe-Salpeter equation (BSE) in the particle-particle channel~ \cite{Maier2006,scalapino2007numerical}. Further details on the DCA and BSE are presented in Secs.~III and IV of SM, respectively.

%\section{Results}

Our main results for the superconducting properties are presented in two parts, based on different cluster size used in the calculations: (A) $N_c$=4$\times$2-site cluster via the BSE; (B) $N_c$=1$\times$2-site cluster via the pair-field susceptibility. Importantly, the $s^\pm$-wave pairing SC is generally enhanced by electron doping, irrespective of the cluster size.

%\subsection{SC with $N_c=4\times2$}

\textbf{\textit{SC with $N_c$=4$\times$2.}} The main result is summarized in the phase diagram shown in Fig.~\ref{phase}. For all three cases, the $s^\pm$-wave pairing SC exhibits a dome-like dependence of $T_c$ on electron doping. Notably, even at 0\,GPa, where no SC exists without doping, a dome-like $T_c$ emerges upon electron doping, peaking around optimal $\delta$$\approx$0.1 and ultimately disappearing near $\delta$$\approx$0.2. For 15\,GPa, the $T_c$ dome is wider, with an optimal doping of $\delta$$\approx$0.2, which agrees with previous slave-boson mean-field results \cite{YF2,YF2026}, where electron doping under pressure also enhances SC. In the LaNi(Al)O case, SC exists without doping and is enhanced by electron doping, with a rapid $T_c$ rise and decay than that of 15\,GPa. Therefore, via electron doping, LaNi(Al)O at ambient pressure can achieve a higher $T_c$ than pressurized La$_3$Ni$_2$O$_7$ (e.g., at $\delta$=0.1 of blue line in Fig.~\ref{phase}). Although the $T_c$ with $N_c$=4 per layer might be overestimated~\cite{Nc4}, the qualitative doping evolution should be trustworthy.

The eigenvectors presented in Fig.~\ref{BSE} (d) and (e) correspond to the lowest accessible temperatures for two representative cases: (d) 15\,GPa with $\delta$=0.3 and (e) LaNi(Al)O with $\delta$=0.1. In both cases, the dominant pairing arises from the $d_{z^2}$ orbital, which is consistent with previous work on a larger $N_c$=16(=8$\times$2)-site cluster using a parameter-reduced version of the model~\cite{maiernpj}. The fact that the interlayer component $\phi_\perp^{zz}$ exceeds the intralayer components $\phi_\|^{zz}$ is a clear signature of $s^\pm$-wave pairing symmetry. Furthermore, in the 15\,GPa case, the pairing components associated with the $d_{x^2-y^2}$ orbital also exhibit a weaker $s^\pm$-wave pairing. This is likely induced by the $d_{z^2}$ orbital through inter-orbital hybridization~\cite{YYF} and/or Hund's coupling~\cite{WCJ2024,YFarxiv,YF2026}. As discussed next, the features are more pronounced in $N_c$=1$\times$2 results at lower temperatures. 

The $T_c$ shown in Fig.~\ref{phase} are obtained by extrapolating the BSE eigenvalues using either a logarithmic function ($\sim A \log(T/T_c)$) or a linear function ($\sim A$\,($T$-$T_c$))~\cite{Maier2019,eigenlog}. More details are given in Sec.~IV of SM~\cite{SM}. As shown in Fig.~\ref{BSE} (a), (b), and (c), although the achievable temperature range is limited by the QMC sign problem, the extrapolated $T_c$ using the appropriate fitting functions can still provide at least qualitatively reasonable doping dependent tendency. Specifically, based on the temperature evolution of the eigenvalues, linear fitting is adopted for all doping levels at 0\,GPa as well as the maximum doping levels of another two cases, i.e., $x$=0.4 at 15\,GPa and $x$=0.2 at LaNi(Al)O. Meanwhile, logarithmic fitting is used for the lower doping levels in the 15\,GPa and LaNi(Al)O cases.  

\begin{figure}
\psfig{figure=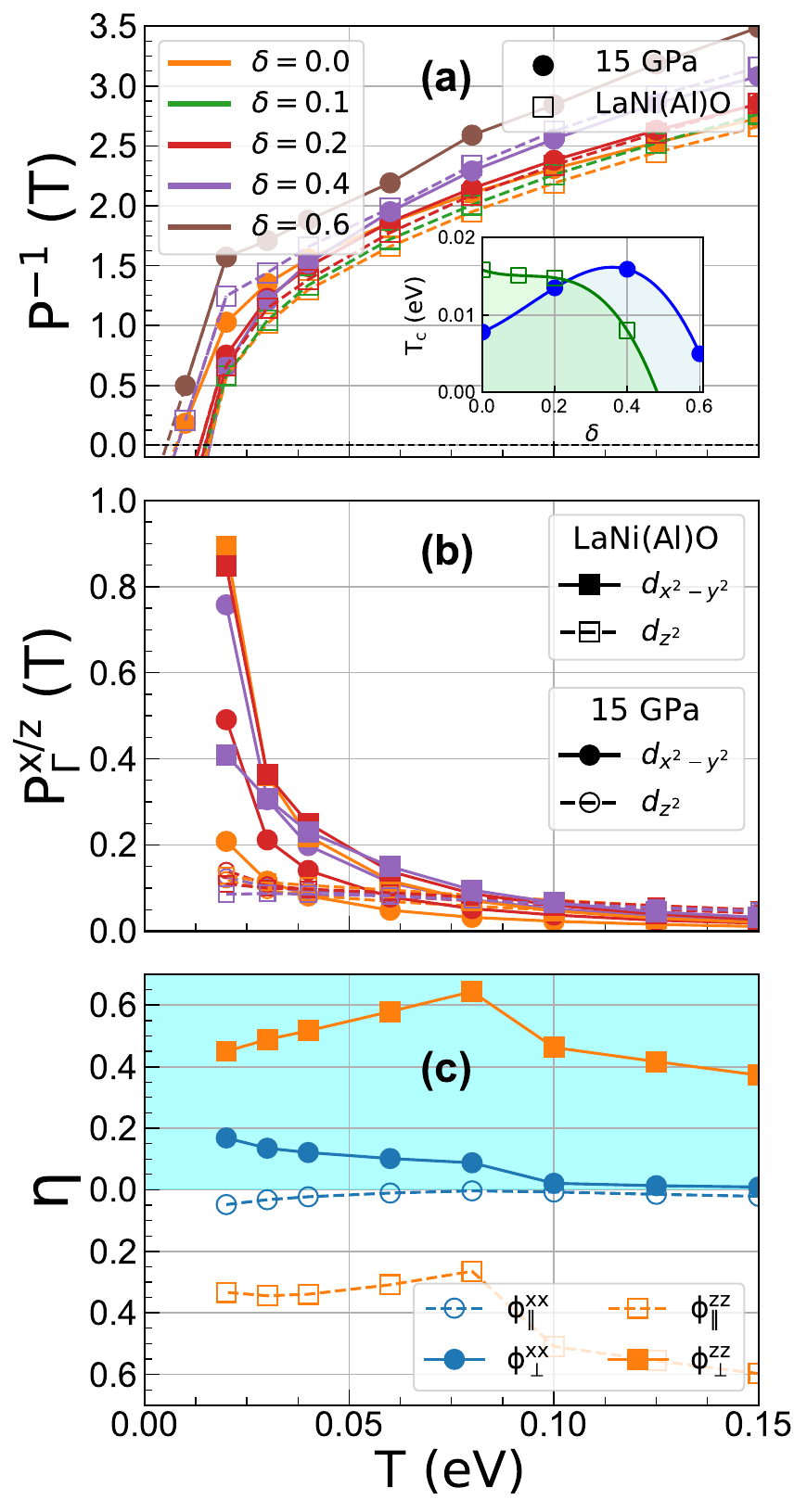,width=0.48\textwidth,clip}
\caption{(a) Temperature dependence of the inverse of $s^\pm$-wave pair-field susceptibility $P^{-1}(T)$ for the 15\,GPa and LaNi(Al)O cases under different electron doping levels $\delta$. The inset shows the corresponding extrapolated $T_c$ obtained from $P^{-1}(T_c)$=0; (b) Orbital-resolved components ($P^x_\Gamma$ and $P^z_\Gamma$) of the interacting part of $s^\pm$-wave pair-field susceptibility at $\delta$=0.0, 0.2, and 0.4 for both cases; (c) Characteristic temperature evolution of the weights $\eta$ (defined in the main text) for the orbital-resolved intralayer ($\phi^{xx/zz}_\|$) and interlayer ($\phi^{xx/zz}_\perp$) eigenvector for the 15\,GPa case at $\delta$=0.2.}
\label{Nc1}
\end{figure}

%\subsection{SC with $N_c=1\times2$}

\textbf{\textit{SC with $N_c$=1$\times$2.}} To explore the superconducting properties at lower temperature scale, we reduce the cluster size to $N_c$=1$\times$2 to examine the orbital-resolved pair-field susceptibility~\cite{maier2001,WW2015,WW2025,maiernpj} for 15\,GPa and LaNi(Al)O cases. 

As shown in Fig.~\ref{Nc1}(a), with the temperature decreasing to $T$=0.01\,eV, the inverse $s^\pm$-wave pair-field susceptibility $P^{-1}(T)$ approaches to zero or becomes negative. For the former, we can extrapolate $T_c$ by linearly fitting $P^{-1}(T)$ at the two lowest temperatures; while for the latter, the negative $P^{-1}(T)$ indicates that $P(T)$ has diverged, which is used to estimate $T_c$. The corresponding $T_c$ is shown in the inset of Fig.~\ref{Nc1}(a). For the 15\,GPa case, a dome-like behavior is observed with a wider electron-doping regime compared to the $T_c$ shown in Fig.~\ref{phase}. For the LaNi(Al)O case, although $T_c$ is also higher than that of the 15\,GPa case at $\delta$=0.0 and 0.1, the dome-like behavior disappears, and instead a half-dome behavior is observed over a wide doping regime. The discrepancy of $T_c$ curves between $N_c$=4$\times$2 and $N_c$=1$\times$2 might originate from the uncertainty of $T_c$ estimation from different temperature ranges, especially for the elevated temperature simulations for $N_c$=4$\times$2. Another possibility is the spatial correlation included in $N_c$=4$\times$2 cluster has apparent impact on the $T_c$ scale and its corresponding doping evolution. In essence, our simulations using two different cluster sizes are complement to each other.

As discussed in detail in Sec.~V of SM~\cite{SM}, the pair-field susceptibility can be decomposed into the non-interacting and interacting parts for both intra- and inter-orbital components. 
Generally, the temperature evolution of the interacting part $P_\Gamma(T)$ is crucial for understanding the correlation induced SC instability~\cite{maiernpj}. Fig.~\ref{Nc1}(b) presents the difference between $P^{x}_\Gamma(T)$ and $P^{z}_\Gamma(T)$ for $\delta = 0.0$, $0.2$, and $0.4$, respectively, in the temperature regime corresponding to the normal state above $T_c$, where the BSE provides a reliable description of superconducting instability. Clearly, as the temperature lowers, $P^{x}_\Gamma(T)$ increases rapidly and eventually surpasses $P^{z}_\Gamma(T)$; while $P^{z}_\Gamma(T)$ remains almost unchanged for both the 15\,GPa and LaNi(Al)O cases, which strongly indicates that the more itinerant $d_{x^2-y^2}$ orbital plays a vital role in the interlayer pairing SC in La$_3$Ni$_2$O$_7$. 

Moreover, although there is no momentum resolution for $N_c$=1$\times$2 cluster, we can explore the representative eigenvector weight $\eta$, which is defined as the absolute value of a given eigenvector component ($\phi^{xx}_{\parallel}$, $\phi^{xx}_{\perp}$, $\phi^{zz}_{\parallel}$, $\phi^{zz}_{\perp}$) at lowest Matsubara frequency normalized by the summation of all four. As a representative example, Fig.~\ref{Nc1}(c) displays the temperature evolution of these weights for the 15\,GPa case at $\delta$=0.2. More results are shown in Sec.~VI of SM~\cite{SM}. Fig.~\ref{Nc1}(c) illustrates that, at relatively high temperature range above $T>0.08$\,eV, the intralayer $\phi^{zz}_{\|}$ of the $d_{z^2}$ orbital decreases while its interlayer counterpart $\phi^{zz}_{\perp}$ increases, signaling a crossover from intralayer to interlayer pairing. However, with further lowering $T$ below $T$=0.08\,eV, the interlayer $\phi^{zz}_{\perp}$ starts to weaken and simultaneously the $d_{x^2-y^2}$ orbital plays a more and more important role before finally exceeding the contribution from $d_{z^2}$ orbital. Apparently, the interlayer $\phi^{xx}_{\perp}$ dominates over the intralayer $\phi^{xx}_{\|}$. Both Fig.~\ref{Nc1}(b-c) suggest that the pairing in the $d_{x^2-y^2}$ orbital becomes more significant than $d_{z^2}$ orbital at low temperature but the pairing may originate from the $d_{z^2}$ orbital although the exact mechanism is unclear yet ~\cite{WCJ2024,YFarxiv,YF2026,YYF,WW327} and deserves more extensive investigation.

%\section{summary and outlook}

\textbf{\textit{Summary and Outlook.}} 
%The current theoretical and experimental studies on nickelate superconductors, particularly those with Ruddlesden-Popper (RP) phases such as La$_3$Ni$_2$O$_7$ and La$_4$Ni$_3$O$_{10}$, predominantly rely on hole doping, for example, by partially substituting La$^{3+}$ with divalent Sr$^{2+}$ \cite{hao2025superconductivity,liu2025temperature}. 
How to achieve electron doping in lanthanum-nickelate superconductors, and the resulting effects on their physical properties, remain key open questions from both experimental and theoretical perspectives.
In the current research, we systematically investigated the superconducting properties of the two-orbital bilayer model for La$_3$Ni$_2$O$_7$ under electron doping, considering La$_3$Ni$_2$O$_7$ at three distinct conditions: bulk La$_3$Ni$_2$O$_7$ at ambient pressure, bulk La$_3$Ni$_2$O$_7$ at 15\,GPa, and a disorder‑free heterostructure La$_3$Ni$_2$O$_7$:La$_3$Al$_2$O$_7$ [denoted LaNi(Al)O] that offers a feasible route to electron doping. 

Using large‑scale dynamical cluster approximation (DCA) simulations with a continuous‑time quantum Monte Carlo cluster solver, and employing cluster sizes $N_c$=4$\times$2 and 1$\times$2, we provide robust numerical evidence that electron doping universally enhances $s^\pm$-wave pairing superconductivity, and $T_c$ exhibits a dome‑like dependence on doping concentration---reminiscent of electron-doped two-dimensional cuprate superconductors \cite{RevModPhys.82.2421}. Notably, the LaNi(Al)O heterostructure, a feasible experimental platform for electron doping, shows the highest $T_c$ in the underdoped regime. Consistent results are also obtained with the smaller cluster size $N_c$=1$\times$2 at lower temperatures, confirming the stability of our conclusions.

Next, the temperature evolution of the orbital‑resolved pair‑field susceptibility and its eigenvector indicates that the $d_{x^2-y^2}$ orbital plays a more dominant role in 
$s^\pm$-wave pairing. Nevertheless, the $s^\pm$-wave pairing on the $d_{x2-y2}$ orbital at low temperatures may be induced by that on the $d_{z^2}$ orbital, implying a two‑orbital cooperative mechanism behind the superconducting instability in  La$_3$Ni$_2$O$_7$---a mechanism that cannot be captured by the single‑band ($d_{x^2-y^2}$) approximation successfully applied to infinite‑layer nickelates \cite{kitatani2020nickelate,PhysRevLett.130.166002,PhysRevResearch.6.043104}. In this two-band model, the inter‑orbital coupling between  $d_{x^2-y^2}$ and $d_{z^2}$ may arise from three possible origins: inter-orbital Hund's coupling \cite{WCJ2024,YFarxiv,YF2026}, inter‑orbital hybridization \cite{YYF}, or a combination of both. The Hund's coupling mechanism is more suitable for integer electron occupancy (i.e., the undoped case), whereas inter‑orbital hybridization becomes relevant for doped systems, consistent with earlier studies on hole‑doped  La$_3$Ni$_2$O$_7$ \cite{WW327}.

Although our many‑body calculations show that electron doping can enhance the superconducting transition temperature, experimental realization remains challenging. The Ce‑doping approach, successful in electron‑doped cuprates \cite{RevModPhys.82.2421}, is ineffective in nickelates because the Ce‑4$f$ electrons lie at much lower energy, making electron hardly transfer into Ni‑$e_g$ orbital. The proposed heterostructure route requires the synthesis of stable La$_3$Ni$_2$O$_7$ layers and the prevention of intermixing between Ni and Al, which also presents significant experimental difficulties. Aside from the heterostructuring way, another possible route to electron doping is hydrogen (H) intercalation. Recent theoretical work \cite{yd8w-frs8} has shown that H-intercalation into La$_2$NiO$_4$ successfully achieves electron doping, resulting in a 3$d^9$ electronic configuration, a single-band $d_{x^2-y^2}$ Fermi surface, an antiferromagnetic Mott insulating ground state---analogous to the parent cuprate La$_2$CuO$_4$---and superconductivity upon hole doping. Hence, (proton) ionic gating (e.g., H$^+$) may provide an alternative way to realize electron doping in bulk and thin film La$_3$Ni$_2$O$_7$, awaiting future experimental verification.

\section{Acknowledgment}

%\textbf{\textit{Acknowledgment.}}
X.~Liu and M.~Jiang acknowledge the support of the National Natural Science Foundation of China (Grant No.12174278) and State Key Laboratory of Surface Physics and Department of Physics in Fudan University (Grant No.KF2025\_12). C.~Deng, W.~Wu, and L.~Si acknowledge support from the National Natural Science Foundation of China (Grant No.~12422407) and the Key Research and Development Program of Shaanxi (2024QY2-GJHX-42). M.~Jiang also acknowledges the support of the China Scholarship Council (CSC) and the hospitality of Karsten Held at TU Wien. Calculations have been mainly done in Soochow University and the National Supercomputing Center (Xi’an) in Northwest University.

\section{DATA AVAILABILITY}

The data that support the findings of this article are not
publicly available. The data are available from the authors
upon reasonable request.

%%%%%%%%%%%%%%%%%%%%%%%%%%%

\clearpage

\bibliography{main}

\end{document}

\title{Supplementary materials to ``Enhanced $s^\pm$-Wave Superconductivity in Electron-Doped La$_3$Ni$_2$O$_7$"}

\author{Xun Liu\,\orcidlink{0009-0000-4906-3078}}
\affiliation{Institute for Quantum Science, School of Physical Science and Technology, \\ Soochow University, Suzhou 215006, China}

\author{Chao Deng\,\orcidlink{0009-0003-8433-2909}}
\affiliation{School of Physics, Northwest University, Xi’an 710127, China}

\author{Wenfeng Wu\,\orcidlink{0000-0002-6575-5813}}
\affiliation{Institute of Solid State Physics, TU Wien, 1040 Vienna, Austria}

\author{Liang Si\,\orcidlink{0000-0003-4709-6882}}
\email{siliang@nwu.edu.cn}
\affiliation{School of Physics, Northwest University, Xi’an 710127, China}
\affiliation{Shaanxi Key Laboratory for Theoretical Physics Frontiers, Xi'an 710127, China}
\affiliation{Fundamental Discipline Research Center for  Quantum Science and technology of Shaanxi Province, Xi'an 710127, China}

\author{Mi Jiang\,\orcidlink{0000-0002-9500-202X}}
\email{jiangmi@suda.edu.cn}
\affiliation{Institute for Quantum Science, School of Physical Science and Technology, \\ Soochow University, Suzhou 215006, China}
\affiliation{State Key Laboratory of Surface Physics and Department of Physics, Fudan University, Shanghai 200433, P. R. China}

\maketitle

\section{DFT calculations}

\begin{figure*}
\psfig{figure=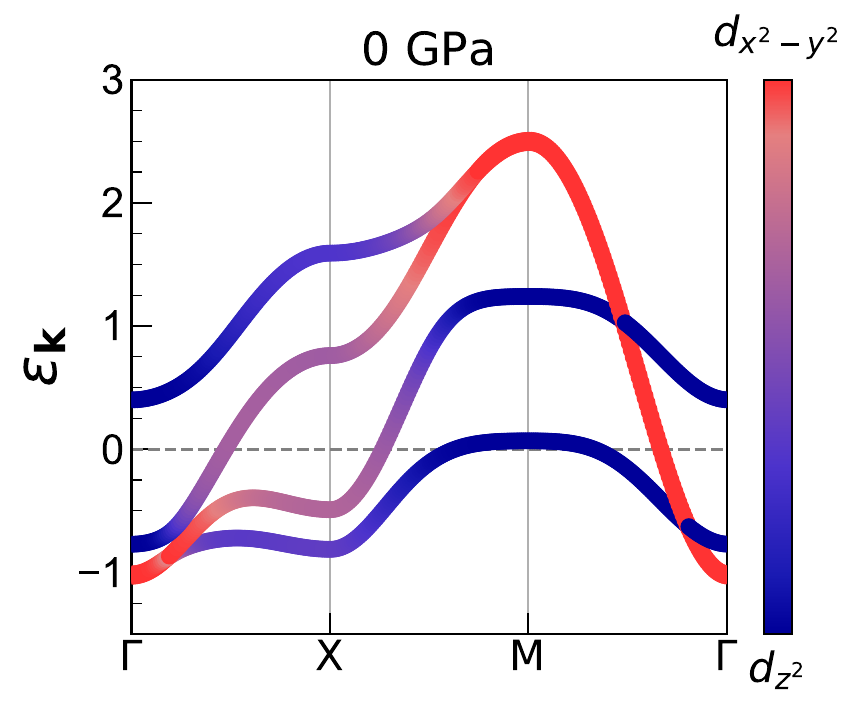,width=0.32\textwidth,clip}
\psfig{figure=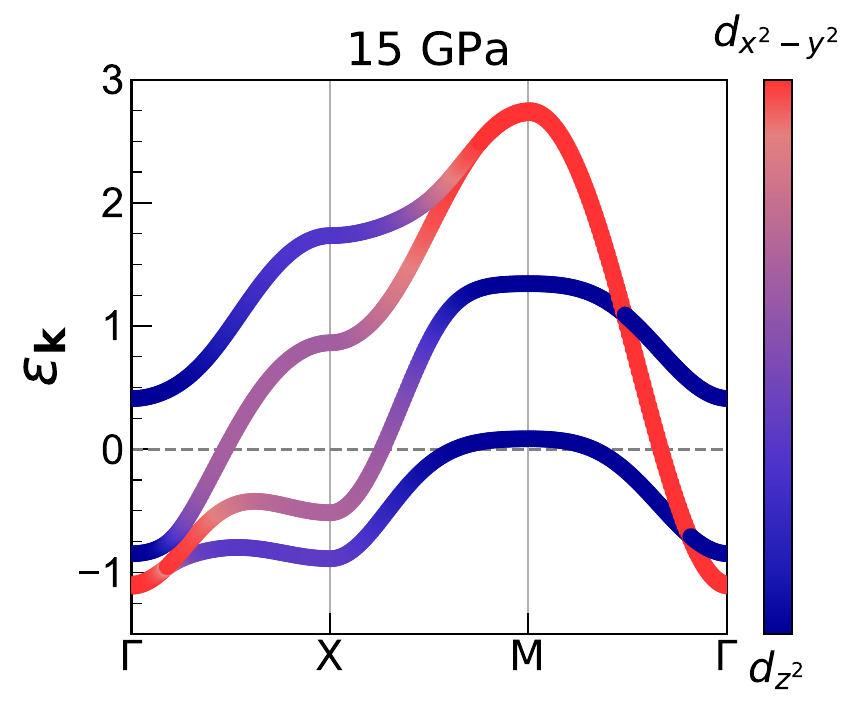,width=0.32\textwidth,clip}
\psfig{figure=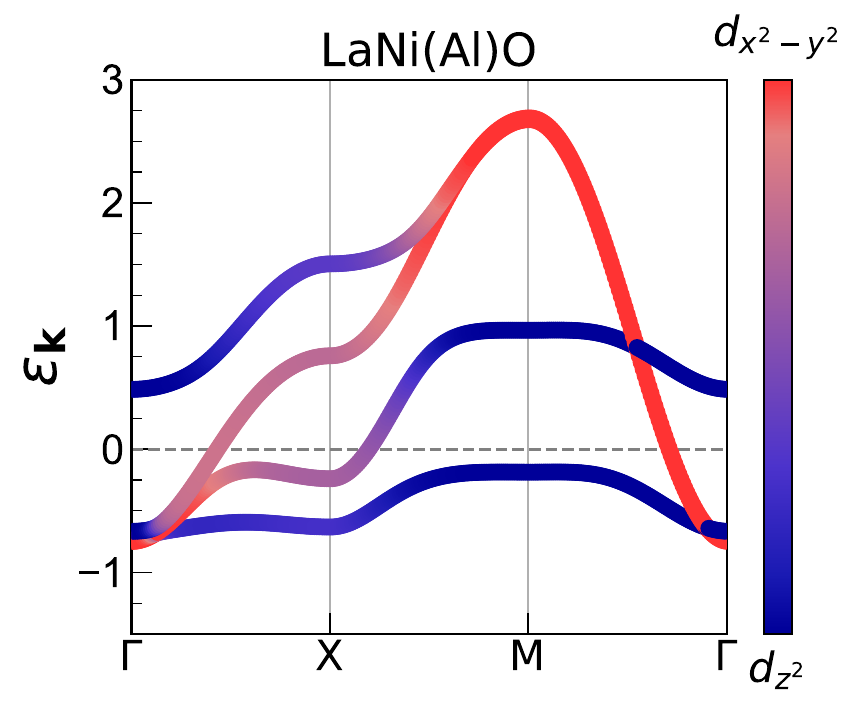,width=0.32\textwidth,clip}
\caption{Non-interacting band structures at $\delta$=0 for three cases: 0\,GPa (left), 15\,GPa (middle), and LaNi(Al)O (right).}
\label{band}
\end{figure*}

The tight-binding parameters shown in Table.~\ref{paras} used in this work are derived from first-principles calculations based on density functional theory (DFT)~\cite{DFT1,DFT2,DFT4,DFT5}. DFT-level structural relaxations and electronic structure calculations were performed using the PAW~\cite{DFT3} method implemented in VASP and the FP-(L)APW method implemented in WIEN2k~\cite{DFT6,DFT7}, with the exchange-correlation functional treated within the GGA-PBE parameterization~\cite{DFT8}. A dense 13×13×3 $k$-point mesh was employed for Brillouin-zone integration of the unit cell. To construct the tight-binding model, the hopping parameters were extracted using maximally localized Wannier functions generated with the Wien2Wannier package~\cite{DFT7}.

The resulting low-energy model retains the Ni-$e_g$ orbitals, $d_{x^2-y^2}$ and $d_{z^2}$, on each Ni site, which dominate the bands near the Fermi level and determine the low-energy Fermi-surface topology of La$_3$Ni$_2$O$_7$. The bilayer two-orbital model therefore provides a widely used minimal description of the relevant low-energy electronic structure and superconducting correlations. We note that the O-$2p$ and other higher-energy orbitals are not treated explicitly; their effects are partially incorporated into the Wannier-derived effective parameters~\cite{YDX}. Although a more complete multi-orbital model may modify quantitative details, it would substantially increase the computational cost and aggravate the fermionic sign problem in the DCA calculations.

For the La$_3$Ni$_2$O$_7$:La$_3$Al$_2$O$_7$ heterostructure, the corresponding parameters were obtained using the same first-principles and Wannier procedure applied to the fully relaxed 1:1 superlattice~\cite{SL}. The La$_3$Al$_2$O$_7$ block acts as an electron donor through interfacial charge transfer, driving the nominal Ni configuration from approximately $3d^{7.5}$ toward $3d^8$. The orbital-resolved DFT electronic structure shows that the low-energy bands remain predominantly of Ni-$e_g$ character, while structural relaxation incorporates the strain and lattice distortions induced by the heterostructure. The resulting four-band tight-binding Hamiltonian, consisting of the $d_{x^2-y^2}$ and $d_{z^2}$ orbitals on the two Ni layers, was used directly in the many-body calculations.

In the DCA calculations, the Ni-$e_g$ occupation obtained from the first-principles electronic structure was used as the reference filling. Electron doping was then modeled by varying the total hole number in the Ni-$e_g$ sector from 2.5 to 2.0, approximately corresponding to the evolution from the $3d^{7.5}$ to the $3d^8$ configuration.

Fig.~\ref{band} shows the non-interacting band structures obtained from DFT parameters in Table~\ref{paras} for the 0\,GPa and 15\,GPa cases, where the bonding ($\gamma$ band) of the $d_{z^2}$ orbital at $M=(\pi,\pi)$ is higher in the 15\,GPa case than in the 0\,GPa case and both cross the Fermi level. In contrast, the $\gamma$ band of LaNi(Al)O lies below the Fermi level, implying larger electron occupation of the $d_{z^2}$ orbital at the same electron density compared to the two former cases.  

\begin{table}
\caption{DFT-derived tight-binding parameters for the bilayer two-orbital model are provided for the following cases: 0\,GPa, 15\,GPa, and the heterostructure La$_3$Ni$_2$O$_7$:La$_3$Al$_2$O$_7$ ~\cite{SL} at 0\,GPa labeled LaNi(Al)O. And the non-interacting band structures are shown in Fig.~\ref{band}. All values are expressed in units of eV.}
\centering
\renewcommand{\arraystretch}{1.2}
\setlength{\extrarowheight}{2pt}
\begin{tabular}{c|ccc}
\hline\hline
\makebox[2.0cm][c]{} & \makebox[2.0cm][c]{0\,GPa} & \makebox[2.0cm][c]{15\,GPa} & \makebox[2.0cm][c]{LaNi(Al)O} \\ [0.8ex]
\hline
\makebox[2.0cm][c]{$\epsilon^x$} & \makebox[2.0cm][c]{0.67145} & \makebox[2.0cm][c]{0.74378} & \makebox[2.0cm][c]{0.36758} \\ [0.8ex]
\hline
\makebox[2.0cm][c]{$\epsilon^z$} & \makebox[2.0cm][c]{0.36885} & \makebox[2.0cm][c]{0.39038} & \makebox[2.0cm][c]{-0.22759} \\ [0.8ex]
\hline
\makebox[2.0cm][c]{$t_1^{xx}$} & \makebox[2.0cm][c]{-0.4400} & \makebox[2.0cm][c]{-0.4804} & \makebox[2.0cm][c]{-0.4279} \\ [0.8ex]
\hline
\makebox[2.0cm][c]{$t_1^{zz}$} & \makebox[2.0cm][c]{-0.1049} & \makebox[2.0cm][c]{-0.1166} & \makebox[2.0cm][c]{-0.0604} \\ [0.8ex]
\hline
\makebox[2.0cm][c]{$t_2^{xx}$} & \makebox[2.0cm][c]{0.0723} & \makebox[2.0cm][c]{0.0755} & \makebox[2.0cm][c]{0.0737} \\ [0.8ex]
\hline
\makebox[2.0cm][c]{$t_2^{zz}$} & \makebox[2.0cm][c]{-0.0162} & \makebox[2.0cm][c]{-0.0169} & \makebox[2.0cm][c]{-0.0211} \\ [0.8ex]
\hline
\makebox[2.0cm][c]{$t_3^{xx}$} & \makebox[2.0cm][c]{-0.0526} & \makebox[2.0cm][c]{-0.0536} & \makebox[2.0cm][c]{-0.0452} \\ [0.8ex]
\hline
\makebox[2.0cm][c]{$t_3^{zz}$} & \makebox[2.0cm][c]{-0.0145} & \makebox[2.0cm][c]{-0.0145} & \makebox[2.0cm][c]{-0.0061} \\ [0.8ex]
\hline
\makebox[2.0cm][c]{$t_\perp^{xx}$} & \makebox[2.0cm][c]{0.00920} & \makebox[2.0cm][c]{0.00815} & \makebox[2.0cm][c]{0.00704} \\ [0.8ex]
\hline
\makebox[2.0cm][c]{$t_\perp^{zz}$} & \makebox[2.0cm][c]{-0.5858} & \makebox[2.0cm][c]{-0.6296} & \makebox[2.0cm][c]{-0.5759} \\ [0.8ex]
\hline
\makebox[2.0cm][c]{$t^{xz}$} & \makebox[2.0cm][c]{0.2154} & \makebox[2.0cm][c]{0.2367} & \makebox[2.0cm][c]{0.1813} \\ [0.8ex]
\hline
\makebox[2.0cm][c]{$t_\perp^{xz}$} & \makebox[2.0cm][c]{-0.025} & \makebox[2.0cm][c]{-0.0257} & \makebox[2.0cm][c]{-0.0270} \\ [0.8ex]
\hline\hline
\end{tabular}
\label{paras}
\end{table}

\section{density distribution}

\begin{figure}
\psfig{figure=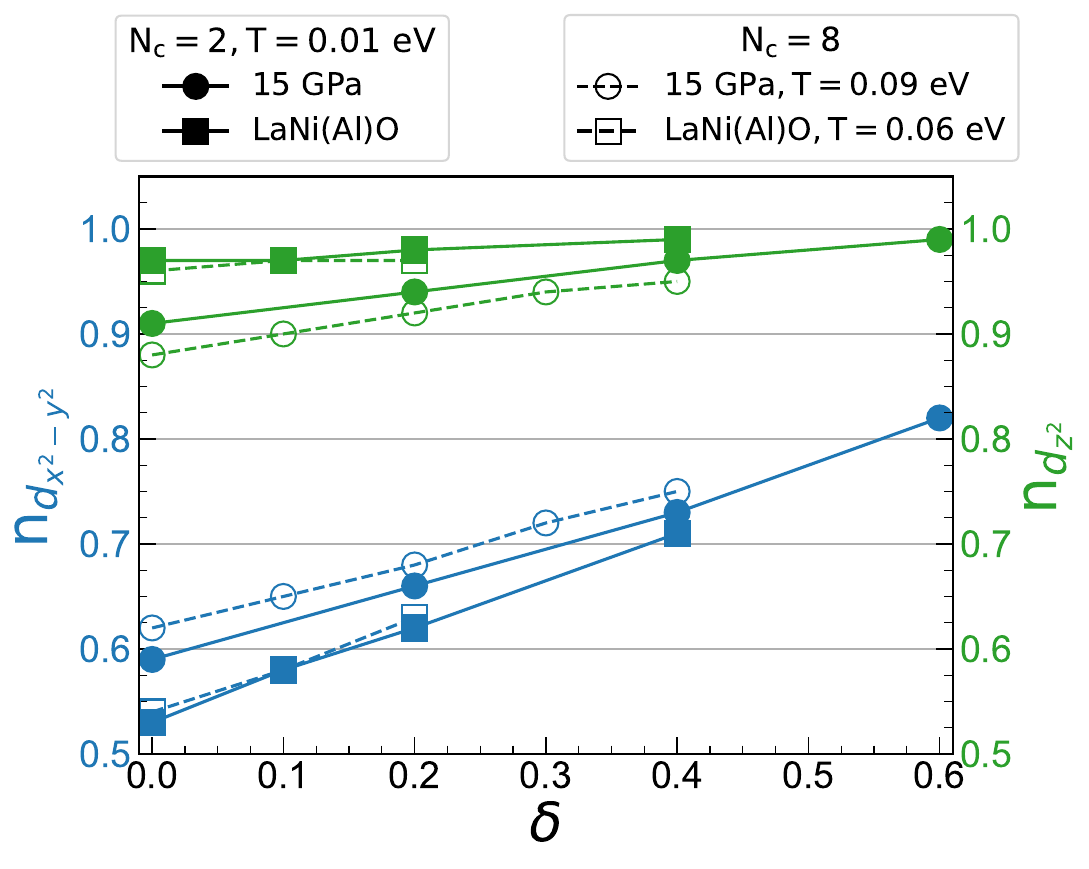,width=0.48\textwidth,clip}
\caption{Electron doping evolution of orbital occupancy for the 15\,GPa (circles) and LaNi(Al)O (squares) cases at the respective lowest achievable temperatures.} 
\label{dens}
\end{figure}

Fig.~\ref{dens} displays the orbital occupancy for the $N_c$=1$\times$2 and $N_c$=4$\times$2 clusters at their lowest accessible temperatures for the 15\,GPa and LaNi(Al)O cases, which confirms the expected density distribution discussed above. For the 15\,GPa case, although the density distributions differ between the two clusters, the orbital occupancies for both orbitals increase with the electron doping level $\delta$, with more doped electrons residing on the $d_{x^2-y^2}$ orbital. In contrast, for LaNi(Al)O, the density of the $d_{z^2}$ orbital ($\approx 0.97$) remains almost unchanged with $\delta$.

\section{Dynamical cluster approximation (DCA)}

The bilayer two-orbital model is solved numerically using the dynamical cluster approximation (DCA) in combination with the continuous-time auxiliary-field (CT-AUX) quantum Monte Carlo (QMC) cluster solver~\cite{Hettler98,Maier05,code,GullCTAUX}. As a celebrated many-body numerical approach, DCA accesses the thermodynamic limit observables by self-consistently embedding a finite cluster in a mean-field bath~\cite{Hettler98,Maier05}. Self-consistency is achieved by iterating between the cluster Green's function and a coarse-grained lattice Green's function, where the coarse-graining is performed over a patch of the Brillouin zone surrounding each cluster momentum $\mathbf{K}$.

Within this framework, short-range correlations inside the cluster are treated exactly using a cluster solver e.g. CT-AUX; while longer-range physics is captured approximately via the mean-field bath. Consequently, enlarging the cluster size systematically approaches the exact thermodynamic limit. The finite cluster effectively discretizes the Brillouin zone into a set of $\mathbf{K}$ points, so that the self-energy $\Sigma(\mathbf{K},i\omega_n)$ is taken as constant within each patch and piecewise constant across the full zone.

\begin{figure}
\psfig{figure=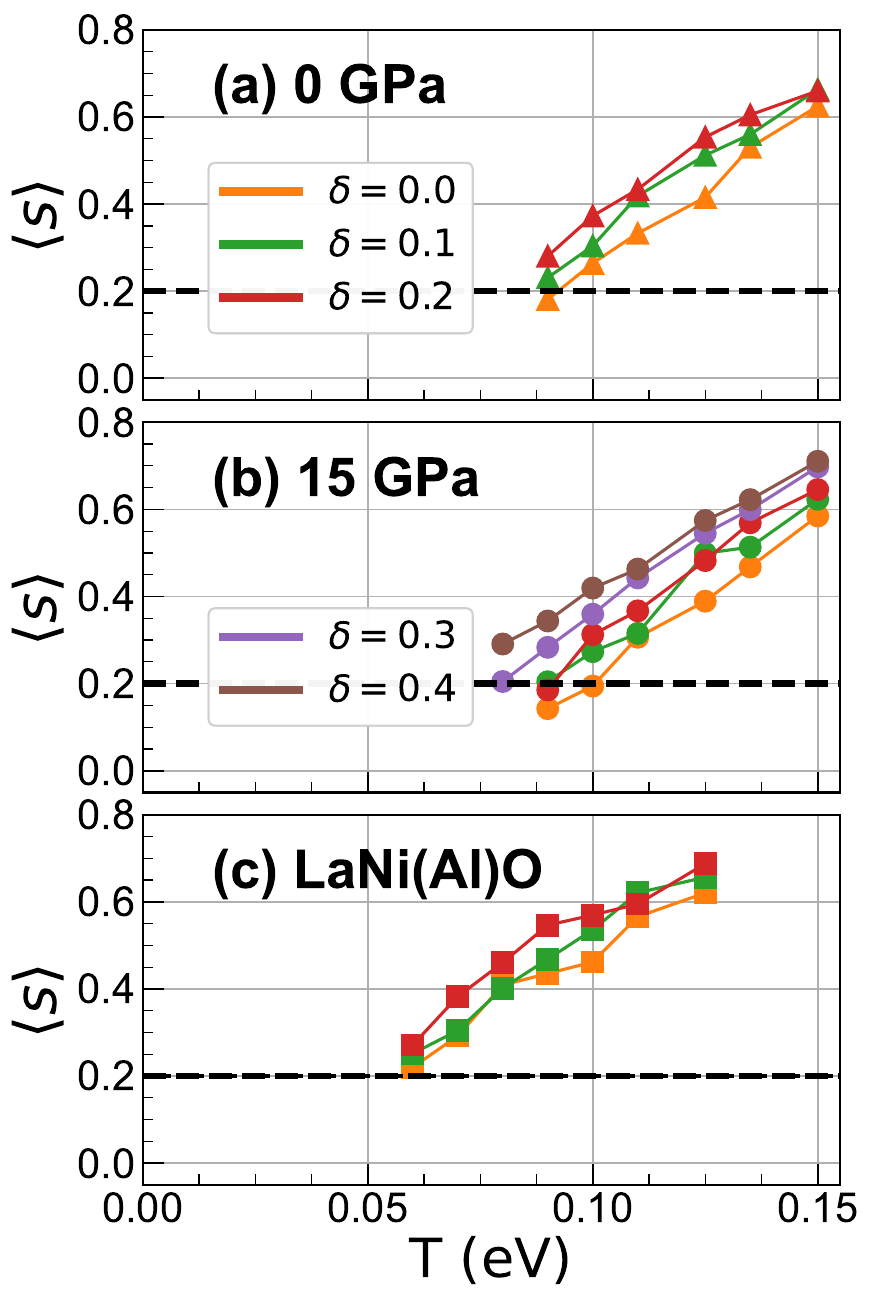,width=0.45\textwidth,clip}
\caption{Temperature dependence of the fermionic sign problem $\langle S \rangle$ on the $N_c$=4$\times$2-site cluster for the three distinct cases: (a) 0\,GPa; (b) 15\,GPa; and (c) LaNi(Al)O. The black dashed line denotes the $\langle s \rangle$=0.2 threshold.
}
\label{Sign}
\end{figure}

Quantum embedding methods like DCA typically suffer from a less severe sign problem than finite-size QMC simulations due to the presence of the mean field. However, owing to model complexity, our calculations are performed on a $N_c$=4$\times$2-site DCA cluster, which includes the $(\pi,0)$ and $(0,\pi)$ points. The corresponding temperature evolution of the sign problem for this cluster is shown in Fig.~\ref{Sign}. The average sign $\langle s \rangle$ indicates that the sign problem becomes more severe at lower temperatures, while it is more moderate for larger doping levels. To ensure the reliability of observables, we impose a threshold $\langle s \rangle \ge 0.2$. In addition, a smaller $N_c$=1$\times$2-site cluster is employed to access lower temperatures, where the sign problem is essentially absent.

\section{Bethe-Salpeter equation (BSE)}

The superconducting properties can be studied by solving the Bethe-Salpeter equation (BSE) in its eigenvalue form in the particle-particle channel~\cite{Maier2006,scalapino2007numerical}
\begin{align} \label{BSE}
-\frac{T}{N_c}\sum_{K'}
\Gamma^{pp}(K,K')
\bar{\chi}_0^{pp}(K')\phi_\alpha(K') =\lambda_\alpha(T) \phi_\alpha(K)
\end{align}
where $\Gamma^{pp}(K,K')$ denotes the irreducible particle-particle vertex of the effective cluster problem, with $K$=($\mathbf{K}, i\omega_n)$ combining the cluster momenta $\mathbf{K}$ and Matsubara frequencies $\omega_n$=$(2n+1)\pi T$, and $\phi_\alpha(K)$ represents the eigenvector obtained by solving the BSE for the pairing symmetry of type $\alpha$.

The coarse-grained bare particle-particle susceptibility
\begin{align}\label{eq:chipp}
	\bar{\chi}^{pp}_0(K) = \frac{N_c}{N}\sum_{k'}G(K+k')G(-K-k')
\end{align}
is obtained via the dressed single-particle Green's function,
\begin{align}
G(k)\equiv G({\bf k},i\omega_n) =
[i\omega_n+\mu-\varepsilon_{\bf k}-\Sigma({\bf K},i\omega_n)]^{-1}
\end{align}
where $\mathbf{k}$ belongs to the DCA patch surrounding the cluster momentum $\mathbf{K}$. with the chemical potential $\mu$ and
\[
\varepsilon_\mathbf{k}
=
\begin{bmatrix}
H_{11}& H_{12} \ \\
H_{21} & H_{22}
\end{bmatrix}
\]
where
\[
H_{11}=H_{22}
=
\begin{bmatrix}
E^{xx} & E^{xz} \ \\
E^{xz} & E^{zz}
\end{bmatrix};H_{12}=H_{21}
=
\begin{bmatrix}
t_\perp^{xx} & E_\perp^{xz} \ \\
E_\perp^{xz} & t_\perp^{zz}
\end{bmatrix}
\]
with
\begin{align}
E^{xx/zz}&=\epsilon^{x/z}+2t_1^{xx/zz}(\cos k_x+\cos k_y) \notag\\
+4&t_2^{xx/zz}\cos k_x\cos k_y +2t_3^{xx/zz}(\cos 2k_x+\cos 2k_y) \notag \\
\notag \\
E^{xz}=2&t^{xz}(\cos k_x-\cos k_y); E_{\perp}^{xz}=2t_\perp^{xz}(\cos k_x-\cos k_y) \notag
\end{align}
and $\Sigma({\mathbf K},i\omega_n)$ the cluster self-energy. In practice, we calculate 32 or more discrete points for both the positive and negative fermionic Matsubara frequency $\omega_n=(2n+1)\pi T$ mesh for measuring the two-particle Green's functions and irreducible vertices. Therefore, the BSE Eq.~\eqref{BSE} reduces to an eigenvalue problem of a matrix of size (128$N_c)\times($128$N_c$). 

%\lx{To characterize pair association, the pair susceptibility is defined as
%\begin{align}
%P_\alpha(T) &= \int_0^l d\tau \langle \Delta_\alpha(\tau) \Delta_\alpha^{\dagger}(0) \rangle 
%\end{align}
The pairing operator is defined as
\begin{align}
\Delta_\alpha^{\dagger} &= \frac{1}{\sqrt{N}} \sum_{\mathbf{K}} g_\alpha(\mathbf{K}) c_{\mathbf{K}}^\dagger c_{-\mathbf{K}}^\dagger
\end{align}
Here, $\alpha$ denotes the pairing symmetry and the $s^\pm$-wave is represented by the form factors $g_{s^\pm}(\mathbf{K})$=$\cos \mathrm{K_z}$.

We analyze the temperature evolution of the BSE eigenvalue $\lambda_\alpha(T)$ ($\alpha$=$s^\pm$). The observation of BCS-like logarithmic temperature dependence would indicate behavior analogous to that observed in the $d$-wave superconducting phase of the conventional single-band Hubbard model in the overdoped regime~\cite{Maier2019,eigenlog}. Conversely, the emergence of linear or exponential temperature dependence would suggest non-BCS pairing fluctuations, similar to those characteristic of the pseudogap regime in the single-band model~\cite{Maier2019}. In practice, we tested both linear and logarithmic fitting within our accessible temperature range. It is indeed possible that the temperature evolution would change at even lower temperature but that is out of the capability of our DCA simulations due to numerical uncertainty.

\begin{figure*}
\psfig{figure=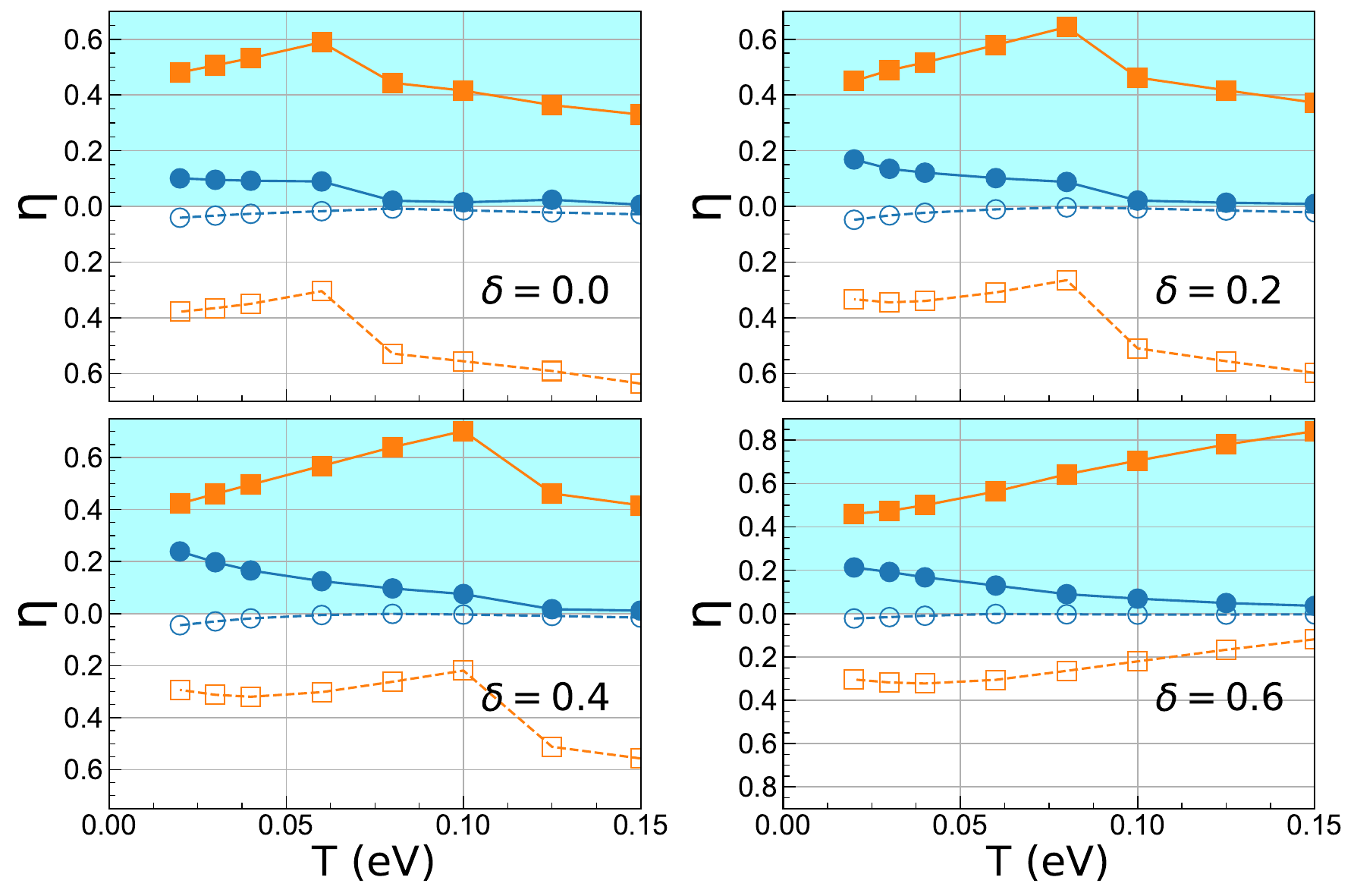,width=0.96\textwidth,clip}
\caption{Eigenvector weights as a function of doping $\delta$=0.0, 0.2, 0.4 and 0.6 for the 15\,GPa case.}
\label{phi1}
\end{figure*}

\begin{figure*}
\psfig{figure=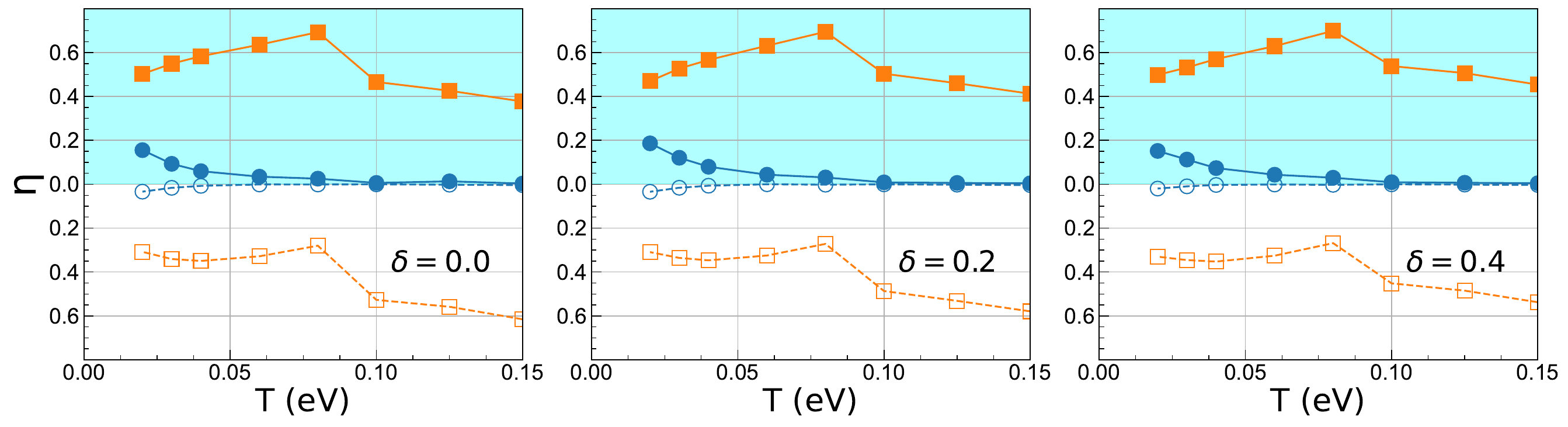,width=0.96\textwidth,clip}
\caption{Akin to Fig.~\ref{phi1}, but for the LaNi(Al)O case. Eigenvector weights as a function of doping $\delta$=0.0, 0.2, and 0.4.}
\label{phi2}
\end{figure*}

\section{$s^\pm$-wave pair-field susceptibility}

Following the usual DCA formalism~\cite{Hettler98,Maier05,code,GullCTAUX}, the pair-field susceptibility is defined as
\begin{align}
P(T)&=\frac{T^2}{N^2_c}\sum_{KK'l_ml'_{m'}} g(\mathbf{K})\bar{G}^{pp}_{l_ml_ml'_{m'}l'_{m'}}(K,K')g(\mathbf{K'})
\end{align}
where $g(\mathbf{K})=\cos k_z$ ($k_z=0$ and $\pi$ for bonding and antibonding combinations) is the form factor for $s^\pm$-wave pairing. $\bar{G}^{pp}_{l_ml_ml'_{m'}l'_{m'}}(K,K')$ is the coarse-grained four-point two-particle Green's function for the orbital components ($l_ml_ml'_{m'}l'_{m'}$, $l_m$ denotes orbital $l$=($d_{x^2-y^2}(x)$, $d_{z^2}(z)$) in $m$(=1,2)-th layer), so that we can decompose the pair-field susceptibility into two orbital-resolved components corresponding to the orbital components ($l_ml_ml'_{m'}l'_{m'}$) for $l$=$l'$=$d_{x^2-y^2}(x)$ or $d_{z^2}(z)$~\cite{WW2015,WW2025}: $P^x$ for $d_{x^2-y^2}$ orbital and $P^z$ for $d_{z^2}$ orbital, in order to examine their respective roles in $s^\pm$-wave pairing. The coarse-grained four-point two-particle Green's function can be evaluated by the BSE

\begin{align}
\bar{G}^{pp}&_{l_ml_ml'_{m'}l'_{m'}}(K,K')= \notag \\
&\bar{G}^{pp,0}_{l_ml_ml'_{m'}l'_{m'}}(K,K')\delta_{K,K'}+\bar{G}^{pp,\Gamma}_{l_ml_ml'_{m'}l'_{m'}}(K,K')
\end{align}

Thus, we define the non-interacting and interacting parts as~\cite{maiernpj}
\begin{align}
P^{x/z}_0(T)&=\frac{T^2}{N^2_c}\sum_{K,K'}\delta_{K,K'} \notag \\
&\sum_{l_ml'_{m'}}g(\mathbf{K})\bar{G}^{pp,0}_{l_ml_ml'_{m'}l'_{m'}}(K,K')g(\mathbf{K'})\\
P^{x/z}_\Gamma(T)&=\frac{T^2}{N^2_c}\sum_{K,K'} \notag\\
&\sum_{l_ml'_{m'}}g(\mathbf{K})\bar{G}^{pp,\Gamma}_{l_ml_ml'_{m'}l'_{m'}}(K,K')g(\mathbf{K'})
\end{align}

In fact, the non-interacting part $P^{x/z}_0(T)$ is the coarse-grained bare particle-particle susceptibility as defined in Eq.(2). The more important interacting part $P^{x/z}_\Gamma(T)$ can be obtained as
\begin{align}
\bar{G}^{pp,\Gamma}&_{l_ml_ml'_{m'}l'_{m'}}(K,K')=  \notag \\
&\frac{T}{N_c}\sum_{K,K'}\sum_{l^1_{m_1}l^2_{m_2}l^3_{m_3}l^4_{m_4}}\bar{G}^{pp,0}_{l_ml_ml^1_{m_1}l^2_{m_2}}(K,K') \notag\\
&\times\Gamma_{l^1_{m_1}l^2_{m_2}l^3_{m_3}l^4_{m_4}}(K,K')\bar{G}^{pp,0}_{l^3_{m_3}l^4_{m_4}l'_{m'}l'_{m'}}(K,K')
\end{align}
where $\Gamma_{l^1_{m_1}l^2_{m_2}l^3_{m_3}l^4_{m_4}}(K,K')$ is the reducible four-point vertex~\cite{maier2001}. To access sufficiently low temperature, we evaluate the pair-field susceptibility using the smallest $N_c$=$1\times2$-site cluster for moderate QMC sign problem.

\section{Further results on the eigenvector weight}

In the main text, the characteristic eigenvector weight shown in Fig.~5(c) suggests an inter-orbital cooperative mechanism underlying the superconducting instability. Specifically, the interlayer pairing hosted by the $d_{x^2-y^2}$ orbital emerges below a crossover temperature ($\approx$ 0.08\,eV) for $\delta$=0.2 in the 15\,GPa case. As shown in Fig.~\ref{phi1} for the 15\,GPa case, this temperature scale increases with electron doping level. However, for the LaNi(Al)O case shown in Fig.~\ref{phi2}, the crossover temperature remains almost unchanged from $\delta$=0.0 to 0.4. These distinct tendencies can be partly understood via the doping dependent $d_{z^2}$ orbital occupancy shown in Fig.~\ref{dens}, where $n_{d_{z^2}}$ gradually increases towards to half-filling with $\delta$ for 15\,GPa case while the LaNi(Al)O case has almost unchanged near half-filling occupancy. In this manner, near-half-filling occupancy promotes the interlayer pairing of the $d_{z^2}$ orbital so that possesses the higher onset temperature for inducing interlayer pairing on the $d_{x^2-y^2}$ orbital. Moreover, a large density close to half-filling on the $d_{z^2}$ orbital is beneficial for the interlayer pairing owing to the strong local moment for the interlayer singlet formation.

%%%%%%%%%%%%%%%%%%%%%

\clearpage

\bibliography{main}